\documentclass[12pt]{article}
\usepackage{amsmath}
\usepackage{graphics,graphicx,xcolor}

\def\be{\begin{equation}}
\def\ee{\end{equation}}
\def\arr{\begin{array}{rll}}
\def\ea{\end{array}}
\def\bea{\begin{eqnarray}}
\def\eea{\end{eqnarray}}

\def\N2{$N{=}2$}

\def\>{\rangle}
\def\<{\langle}
\def\+{\dagger}
\def\={\ =\ }

\def\bal{\begin{aligned}}
\def\eal{\end{aligned}}
\begin{document}
\begin{titlepage}
\setcounter{page}{0}
\begin{center}
{\LARGE\bf  Generalised point vortices on a plane}\\
\vskip 1.5cm
\textrm{\Large Anton Galajinsky \ }
\vskip 0.7cm
{\it
Tomsk Polytechnic University, 634050 Tomsk, Lenin Ave. 30, Russia
} \\
\vskip 0.1cm
{E-mail: galajin@tpu.ru}
\vskip 0.5cm
\end{center}

\begin{abstract} \noindent
A three--vortex system on a plane is known to be minimally superintegrable in the Liouville sense.
In this work, integrable generalisations of the three--vortex planar model, which involve root vectors of simple Lie algebras, are proposed. It is shown that a generalised system, which is governed by a positive definite Hamiltonian, admits a natural integrable extension by spin degrees of freedom.
It is emphasised that the $n$--vortex planar model and plenty of its generalisations enjoy the nonrelativistic scale invariance, which gives room for possible holographic applications.

\end{abstract}

\vspace{0.5cm}

\end{titlepage}
\renewcommand{\thefootnote}{\arabic{footnote}}
\setcounter{footnote}0

\noindent
{\bf 1. Introduction}\\

\noindent
In the family of integrable models with finite number of degrees of freedom, a system of three point vortices on a plane holds a special place with impressive history. The research began with the work of Helmholtz \cite{HH}, who had demonstrated that $2d$ Euler equations for an incompressible inviscid fluid admitted a particular solution which described $n$
point vortices on a plane. A few decades later, Kirchhoff had rewritten the $n$--vortex equations in the Hamiltonian form (see Lecture 20 in \cite{GK}). First results on integrability of the three--vortex case
were reported by W. Gr\"obli \cite{WG} and Poincar\'e \cite{HP}. The nonintegrability of a generic four--vortex model had been proven almost a century later \cite{Z} (see also a related work \cite{BE}). In modern times, the study of non--planar generalisations and related topics generated extensive literature (see e.g. \cite{BBM,BM,HA,PKN,MV} and references therein).

The integrability of the three--vortex planar model relies upon the $E(2)$--symmetry. In particular, the generator of rotation on a two--dimensional plane, the Casimir element of $e(2)$ and the Hamiltonian itself provide three functionally independent integrals of motion in involution. Adding the generator of translation in one of two spatial directions, renders the system minimally superintegrable.\footnote{Recall that a Hamiltonian system with $2n$ phase space degrees of freedom is called Liouville integrable, if it admits $n$ functionally
independent first integrals, which commute under the Poisson bracket. If there are more than $n$ such integrals, a model is called superintegrable. Because in unparameterised form one has $2n-1$ equations of motion, the maximum possible number of functionally independent first integrals is $2n-1$. A dynamical system possessing $2n-1$ first integrals is called maximally
superintegrable, while that admitting $n+1$ first integrals is named minimally superintegrable.}

The goal of this Letter is threefold. Firstly, it is emphasised that the $n$--vortex planar model and plenty of its generalisations enjoy the nonrelativistic scale invariance. Despite extensive recent studies of the fluid/gravity correspondence (see e.g. \cite{MR} and references therein) the $n$--vortex system appears to have escaped attention. Secondly, it is demonstrated that the $E(2)$--symmetry alone does not fix an integrable Hamiltonian, but rather specifies its arguments. Generalisations involving root vectors of simple Lie algebras are proposed.
Thirdly, it is shown that a generalised three--vortex model, which is governed by a positive definite integrable Hamiltonian, admits a natural integrable extension by spin degrees of freedom.

The work is organised as follows.

In the next section, a system of $n$ point vortices on a plane is reviewed and its invariance under the nonrelativistic scale transformation is established.
In Sect. 3, restrictions on a form of a Hamiltonian, which follow from the $E(2)$--symmetry and the related integrability, are formulated. A few generalisations, which rely upon root vectors of simple Lie algebras, are proposed, some of them bearing resemblance to the Ruijsenaars--Schneider model \cite{RS}. In Sect. 4, it is shown that a generalised
three--vortex system, which is governed by a positive definite  Hamiltonian, can be extended by dynamical spin variables without destroying integrability. In the concluding Sect. 5, we summarise our results and discuss possible further developments.

\vspace{0.5cm}

\noindent
{\bf 2. A system of $n$ point vortices on a plane }\\

\noindent
As is known since Kirchhoff's work (see Lecture 20 in \cite{GK}), a system of $n$ point vortices on a plane can be described by canonical equations of motion which derive from the Hamiltonian (in what follows we use the notation in \cite{K})
\be\label{H}
H=\frac{1}{\pi} \sum_{i\ne j} \Gamma_i \Gamma_j \ln{\left({\left(x_i-x_j\right)}^2+{\left(y_i-y_j \right)}^2 \right)},
\ee
where $(x_i,y_i)$, $i=1,\dots,n$, are Cartesian coordinates of the $i$--th vortex and $\Gamma_i$ is its (constant) circulation,
and the Poisson bracket
\be\label{PB}
\{A,B\}=\sum_{i=1}^n \frac{1}{\Gamma_i} \left(\frac{\partial A}{\partial y_i} \frac{\partial B}{\partial x_i} -\frac{\partial A}{\partial x_i} \frac{\partial B}{\partial y_i} \right).
\ee

The invariance of (\ref{H}) under translations and rotation on a two--dimensional plane results in three constants of the motion
\be\label{IM}
P_x=\sum_{i=1}^n \Gamma_i x_i, \qquad P_y=\sum_{i=1}^n \Gamma_i y_i, \qquad M=\frac 12 \sum_{i=1}^n \Gamma_i \left(x_i^2+y_i^2 \right),
\ee
which obey the structure relations of the (centrally extended) Lie algebra associated with the Euclidean group $E(2)$
\be\label{SR}
\{M,P_x \}=P_y, \qquad \{M,P_y \}=-P_x, \qquad \{P_x,P_y \}=-\sum_{i=1}^n \Gamma_i.
\ee
As follows from (\ref{IM}), (\ref{SR}), the three--vortex case is minimally superintegrable in the Liouville sense. Indeed,
the quadratic combination $P_x^2+P_y^2$ along with $H$ and $M$ provide three functionally independent first integrals in involution, while adding $P_x$ (or $P_y$) renders the system minimally superintegrable.
Note that, if the sum of circulations vanishes, $P_x^2+P_y^2$ coincides with the Casimir element of $e(2)$.

The equations of motion resulting from (\ref{H}), (\ref{PB})
\bea\label{EOM}
&&
{\dot x}_a=-\frac{4}{\pi} \sum_{i \ne a} \frac{\Gamma_i (y_a-y_i)}{{\left(x_a-x_i\right)}^2+{\left(y_a-y_i \right)}^2}, \qquad {\dot y}_a=\frac{4}{\pi} \sum_{i \ne a} \frac{\Gamma_i (x_a-x_i)}{{\left(x_a-x_i\right)}^2+{\left(y_a-y_i \right)}^2},
\eea
also hold invariant under the scale transformation
\be\label{Dil}
x'_i=\lambda x_i, \qquad  y'_i=\lambda y_i, \qquad t'=\lambda^2 t,
\ee
where $\lambda$ is an arbitrary real parameter, which coincides with the dilatation transformation entering the Schr\"odinger group \cite{N}. Because the action functional
\be
S=\int dt \left(\sum_{i=1}^n \Gamma_i x_i {\dot y}_i-H \right)
\ee
associated with Eqs. (\ref{EOM}) transforms as $S'=\lambda^2 S+\mbox{const}$ under the dilatation transformation (\ref{Dil}), the construction of a conserved charge via Noether's theorem appears problematic.
This is also seen from a natural candidate for the dilatation generator $\sum_{i=1}^n x_i y_i$, which fails to produce the infinitesimal form of (\ref{Dil}) via the Poisson bracket (\ref{PB}). To the best of our knowledge, the scaling symmetry of (\ref{EOM}) escaped attention and a gravity dual to a system of $n$ point vortices on a plane has not yet been explored in the literature.

It is worth mentioning that, according to the Jacobi last multiplier method (see e.g. \cite{W}), a system of first--order differential equations ${\dot z}_i=f_i (z)$, $i=1,\dots,m+1$, is integrable by quadratures, if it possesses $m-1$ functionally independent first integrals and admits an integrating multiplier $\mu$ obeying
\be
\dot\mu+\mu \partial_i f_i=0.
\ee
In particular, if a vector field $f_i$ is divergence--free, a system automatically admits an integrating multiplier $\mu=\mbox{const}$. As $\partial_i f_i=0$ for the equations (\ref{EOM}), the three--vortex case can alternatively be studied by applying the Jacobi approach.

\begin{figure}[ht]
\begin{center}
\resizebox{0.5\textwidth}{!}{%
\includegraphics{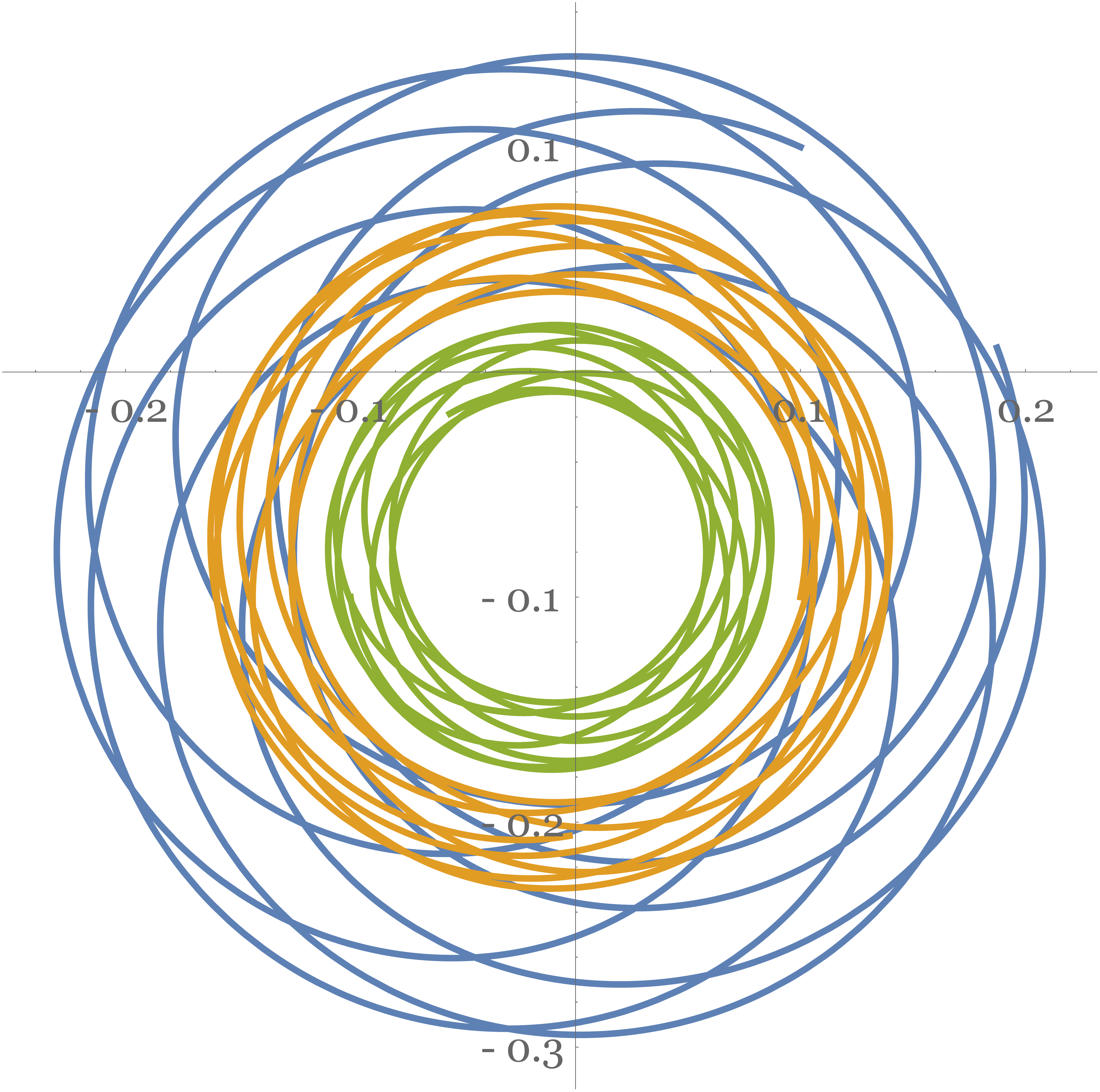}}\vskip-4mm
\caption{\small Parametric plots $(x_1 (t),y_1 (t))$ (outer), $(x_2 (t),y_2 (t))$ (middle), $(x_3 (t),y_3 (t))$ (inner) associated with the Hamiltonian (\ref{H}) for $\Gamma_1=0.1$, $\Gamma_2=0.3$, $\Gamma_3=0.5$, $x_1(0)=0.1$, $y_1 (0)=0.1$, $x_2 (0)=0.1$, $y_2 (0)=-0.1$, $x_3 (0)=-0.1$, $y_3 (0)=-0.1$, and $t \in [0,2]$.}
\label{fig3}
\end{center}
\end{figure}

For what follows, it proves instructive to display parametric plots $(x_1 (t),y_1 (t))$ (outer), $(x_2 (t),y_2 (t))$ (middle), $(x_3 (t),y_3 (t))$ (inner) associated with the Hamiltonian (\ref{H}) for $\Gamma_1=0.1$, $\Gamma_2=0.3$, $\Gamma_3=0.5$, $x_1(0)=0.1$, $y_1 (0)=0.1$, $x_2 (0)=0.1$, $y_2 (0)=-0.1$, $x_3 (0)=-0.1$, $y_3 (0)=-0.1$, and $t \in [0,2]$ (see Fig. 1).

As was mentioned in the Introduction, for $n>3$ and generic values of the circulations $\Gamma_i$ the equations (\ref{EOM}) cease to be integrable \cite{Z}. When discussing generalised models below, we mainly focus on the three--vortex case.
\vspace{0.5cm}

\noindent
{\bf 3.  Generalised three--vortex systems on a plane}\\

\noindent
Eqs. (\ref{EOM}) were originally obtained by invoking basic principles of nonrelativistic fluid mechanics. In particular, specific boundary conditions on a fluid, in which point vortices propagate, were assumed \cite{GK}. In this section, we temporarily set aside physical grounds and bring to the forefront the issues of symmetry and integrability, thus paving the way for generalisations.

Consider an arbitrary function $H(x,y)$ of $(x_i,y_i)$, $i=1,2,3$, which will be identified below with the Hamiltonian of a generalised three--vortex system on a plane, and let us demand it be inert under the action of the Euclidean group $E(2)$ generated by (\ref{IM}) via the Poisson bracket (\ref{PB}). From $\{P_x,H \}=0$, $\{P_y,H \}=0$, $\{M,H\}=0$ one obtains the linear homogeneous partial differential equations
\bea\label{SE}
\sum_{i=1}^3 \frac{\partial H}{\partial y_i}=0, \qquad \sum_{i=1}^3 \frac{\partial H}{\partial x_i}=0, \qquad \sum_{i=1}^3 \left(y_i \frac{\partial H}{\partial x_i}- x_i \frac{\partial H}{\partial y_i} \right)=0,
\eea
which can be solved by the well known method of characteristics. The general solution to the first two equations in (\ref{SE}) is an arbitrary function of the arguments
\be\label{SE1}
x_1-x_2, \qquad x_1-x_3, \qquad y_1-y_2, \qquad y_1-y_3,
\ee
$x_2-x_3$ and $y_2-y_3$ being the linear combinations of the above,
while the ordinary differential equations associated with the third restriction in (\ref{SE})
\be
\frac{dx_1}{y_1}=\frac{dx_2}{y_2}=\frac{dx_3}{y_3}=-\frac{dy_1}{x_1}=-\frac{dy_2}{x_2}=-\frac{dy_3}{x_3}
\ee
give rise to the first integrals
\be\label{SE2}
x_1^2+y_1^2, \qquad x_2^2+y_2^2, \qquad x_3^2+y_3^2, \qquad x_1 x_2+y_1 y_2, \qquad x_1 x_3+y_1 y_3.
\ee
A way to consistently combine (\ref{SE1}) and (\ref{SE2}) is to choose the quadratic combination
\be\label{ARG}
{\left(\alpha (x_i-x_j)+\beta (x_k-x_l)\right)}^2 + {\left(\alpha (y_i-y_j)+\beta (y_k-y_l)\right)}^2,
\ee
where $\alpha$ and $\beta$ are arbitrary real parameters and $i,j,k,l=1,2,3$. The latter features the argument of $H$ obeying (\ref{SE}).

Eq. (\ref{ARG}) allows one to construct a plethora of generalised three--vortex systems on a plane, which are $E(2)$--invariant and minimally superintegrable. For example, regarding the original model (\ref{H}) as being associated with root vectors of the simple Lie algebra $A_2$, and switching instead to long root vectors of $G_2$, one gets the Hamiltonian
\be\label{G2}
H=\frac{1}{\pi} \sum_{i\ne j \ne k} \Gamma_i \Gamma_j \ln{\left({\left(x_i+x_j-2 x_k\right)}^2+{\left(y_i+y_j-2 y_k \right)}^2 \right)},
\ee
which results in more fancy orbits (see Fig. 2).

\begin{figure}[ht]
\begin{center}
\resizebox{0.5\textwidth}{!}{%
\includegraphics{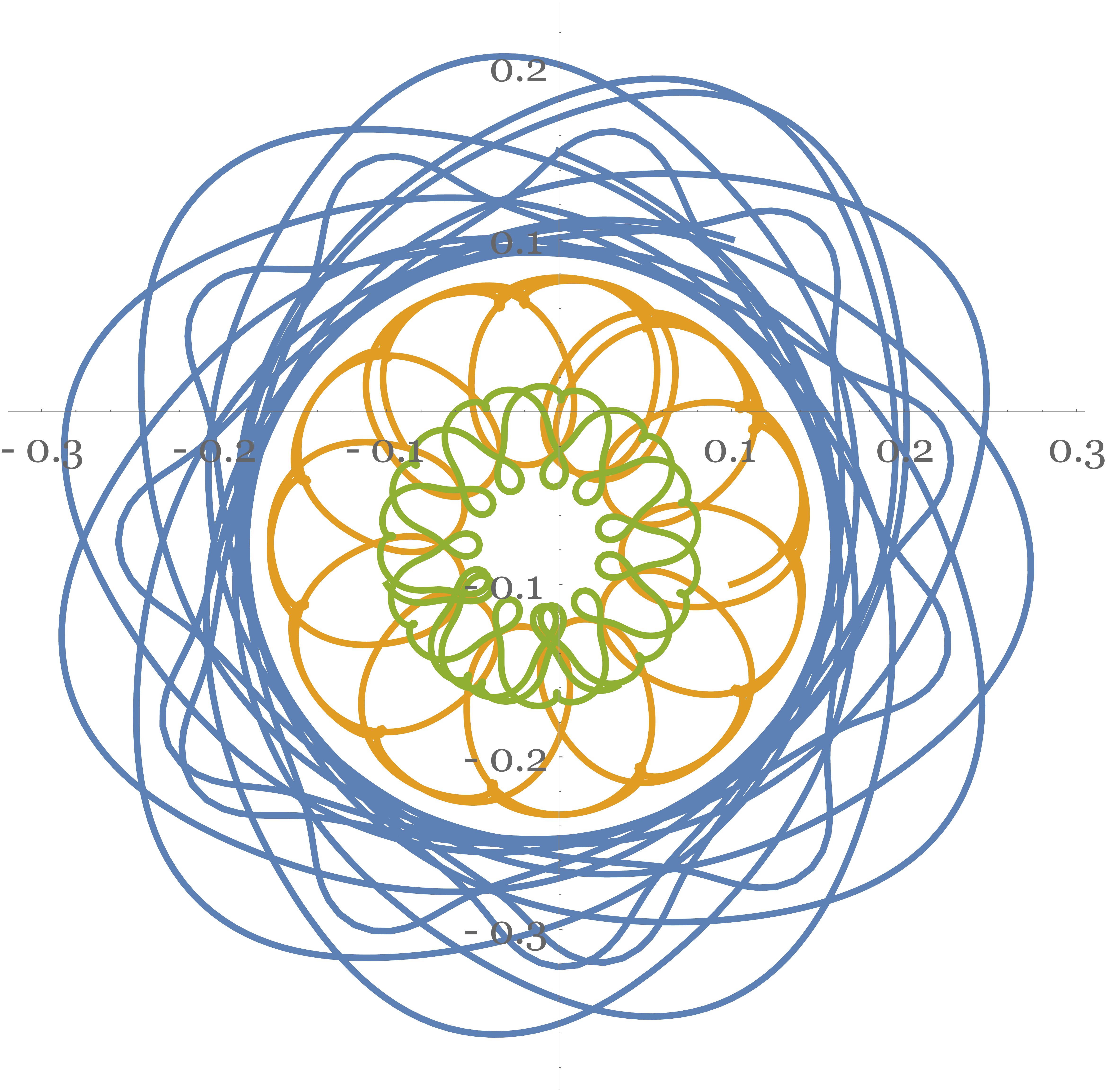}}\vskip-4mm
\caption{\small Parametric plots $(x_1 (t),y_1 (t))$ (outer), $(x_2 (t),y_2 (t))$ (middle), $(x_3 (t),y_3 (t))$ (inner) associated with the Hamiltonian (\ref{G2}), for $\Gamma_1=0.1$, $\Gamma_2=0.3$, $\Gamma_3=0.5$, $x_1(0)=0.1$, $y_1 (0)=0.1$, $x_2 (0)=0.1$, $y_2 (0)=-0.1$, $x_3 (0)=-0.1$, $y_3 (0)=-0.1$, and $t \in [0,2]$.}
\label{fig3}
\end{center}
\end{figure}

Combining (\ref{H}) and (\ref{G2})
\bea\label{G2full}
&&
H=\frac{1}{\pi} \sum_{i\ne j}^3 \Gamma_i \Gamma_j \ln{\left({\left(x_i-x_j\right)}^2+{\left(y_i-y_j \right)}^2 \right)}
\nonumber\\[2pt]
&&
\qquad
+\frac{1}{\pi} \sum_{i\ne j \ne k}^3 W_i W_j \ln{\left({\left(x_i+x_j-2 x_k\right)}^2+{\left(y_i+y_j-2 y_k \right)}^2 \right)},
\eea
where $W_i$ are arbitrary parameters (coupling constants), one gets what can be called a $G_2$ three--vortex system on a plane. In particular, by adjusting the parameters $\Gamma_i$ and $W_i$, one can interpolate between the orbits exposed in Fig. 1 and Fig. 2 Note that, similarly to (\ref{EOM}), the $G_2$--system is invariant under the scale transformation (\ref{Dil}).

As follows from our consideration above, an explicit form of a Hamiltonian is not fixed by demanding the $E(2)$ symmetry alone. So one is at liberty to experiment with various functions of the argument (\ref{ARG}) and build a plethora of generalised systems. An interesting model arises if one replaces the logarithm in (\ref{H}) with the exponent
\be\label{H2}
H=\frac 14 \sum_{i\ne j}\Gamma_i \Gamma_j e^{2 {\left(x_i-x_j\right)}^2} e^{2 {\left(y_i-y_j \right)}^2}.
\ee
Eq. (\ref{H2}) bears resemblance to the Ruijsenaars--Schneider model \cite{RS} and it is characterised by a more gentle dynamical behaviour (see Fig. 3). Note though that it does not enjoy the scale symmetry (\ref{Dil}).

\begin{figure}[ht]
\begin{center}
\resizebox{0.5\textwidth}{!}{%
\includegraphics{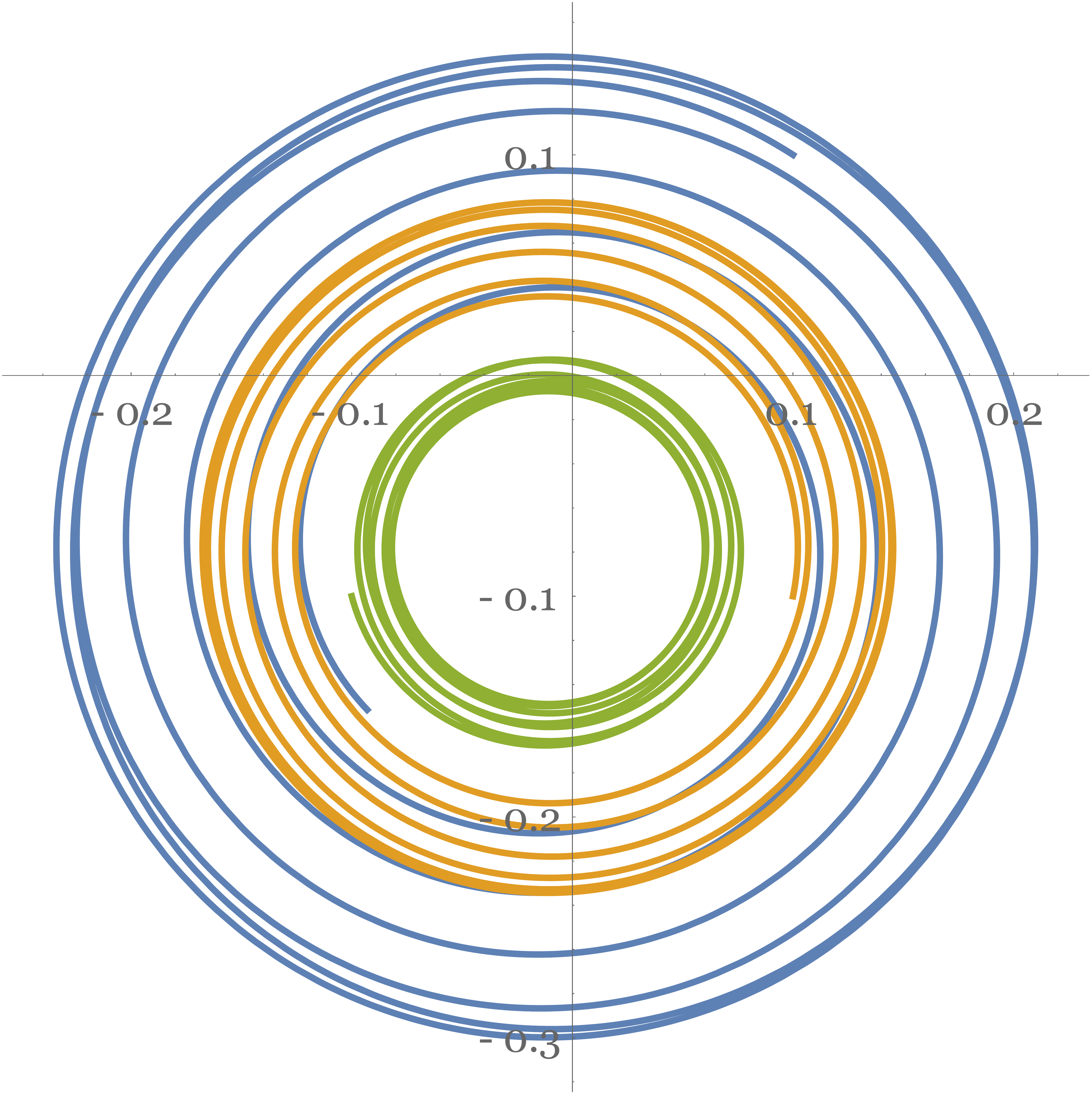}}\vskip-4mm
\caption{\small Parametric plots $(x_1 (t),y_1 (t))$ (outer), $(x_2 (t),y_2 (t))$ (middle), $(x_3 (t),y_3 (t))$ (inner) associated with the Hamiltonian (\ref{H2}), for $\Gamma_1=0.1$, $\Gamma_2=0.3$, $\Gamma_3=0.5$, $x_1(0)=0.1$, $y_1 (0)=0.1$, $x_2 (0)=0.1$, $y_2 (0)=-0.1$, $x_3 (0)=-0.1$, $y_3 (0)=-0.1$, and $t \in [0,20]$.}
\label{fig3}
\end{center}
\end{figure}

It is natural to expect that $n>3$ generalised systems will lack the integrability property. In particular, one can look into the simplest model 
\be\label{H4}
H=\frac{1}{16 \pi} \sum_{i\ne j \ne k \ne l}(\Gamma_i+\Gamma_j)(\Gamma_k+\Gamma_l) \ln{\left( {(x_i+x_j-x_k-x_l)}^2+{(y_i+y_j-y_k-y_l)}^2 \right)},
\ee
and reveal rather erratic orbits depicted in Fig. 4. 

\begin{figure}[ht]
\begin{center}
\resizebox{0.5\textwidth}{!}{%
\includegraphics{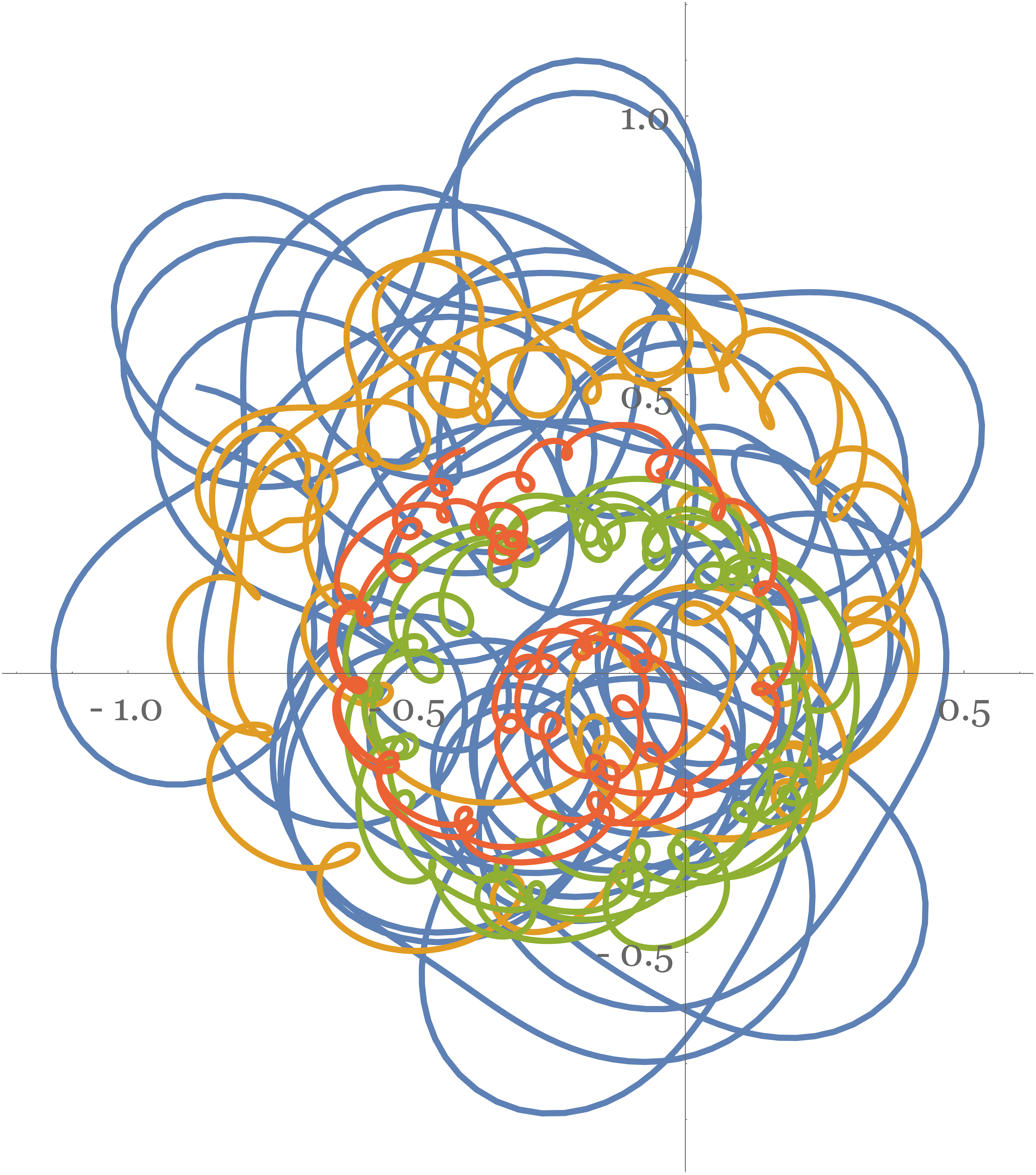}}\vskip-4mm
\caption{\small The orbits associated with the Hamiltonian (\ref{H4}) exhibit erratic behaviour. The plot is given for $\Gamma_1=0.1$, $\Gamma_2=0.3$, $\Gamma_3=0.5$, $\Gamma_4=0.7$, $x_1(0)=0.1$, $y_1 (0)=0.1$, $x_2 (0)=0.2$, $y_2 (0)=-0.2$, $x_3 (0)=-0.3$, $y_3 (0)=-0.3$, $x_4 (0)=-0.4$, $y_4 (0)=0.4$, and $t \in [0,3]$.}
\label{fig3}
\end{center}
\end{figure}

In the next section, we dwell on positive definite Hamiltonians similar to (\ref{H2}) and construct integrable extensions of such systems by spin degrees of freedom.

\vspace{0.5cm}

\noindent
{\bf 4. Integrable extensions by supersymmetrisation}\\

\noindent
As was advocated in a recent work \cite{AG}, given an integrable system with a positive definite Hamiltonian, one can use methods of supersymmetry in order to construct its integrable (bosonic) extension. In this section, we elaborate on this proposal by focusing on generalised three--vortex models.

Let us consider a three--vortex system governed by a positive--definite Hamiltonian
\be\label{HHam}
H=\frac 12 {\left(\Lambda_{12}\right)}^2+\frac 12 {\left(\Lambda_{13}\right)}^2+\frac 12 {\left(\Lambda_{23}\right)}^2,
\ee
where $\Lambda_{ij}$ are three functions of $(x_i,y_i)$, $i,j=1,2,3$, and it is assumed that their arguments are structured in accord with the prescription (\ref{ARG}).
In order to construct an $\mathcal{N}=1$ supersymmetric extension of (\ref{HHam}), for each bosonic pair $(x_i,y_i)$ one introduces a real fermionic partner $\theta_i$, $i=1,2,3$, obeying the Poisson brackets
\be\label{PPbr}
\{\theta_i,\theta_j \}=-{\rm i} \delta_{ij}, \qquad \{\theta_i,x_j \}=0, \qquad \{\theta_i,y_j \}=0,
\ee
and then builds the supersymmetry charge
\be\label{sc}
Q=\Lambda_{12} \theta_3+\Lambda_{13} \theta_2+\Lambda_{23} \theta_1,
\ee
which via the Poisson bracket
\be
\{Q,Q \}=-2 {\rm i} \mathcal{H},
\ee
gives rise to the super--extended Hamiltonian
\be\label{SHam}
\mathcal{H}=H-{\rm i} \{\Lambda_{13},\Lambda_{23} \} \theta_1 \theta_2-{\rm i} \{\Lambda_{12},\Lambda_{23} \} \theta_1 \theta_3-{\rm i} \{\Lambda_{12},\Lambda_{13} \} \theta_2 \theta_3.
\ee
From (\ref{SHam}) one can readily obtain equations of motion describing the extended system. In particular, the original equations ${\dot x}_i=\{x_i,H\}$ and ${\dot y}_i=\{y_i,H\}$ will be modified to include fermionic contributions.

In general, the super-extended system is not integrable, as one introduces three fermionic degrees of freedom and only one conserved super--charge. Yet, one can achieve an integrable generalisation, if one focuses on a {\it particular} solution \cite{AG}, in which all fermions are proportional to one and the same Grassmann--odd number $\epsilon$
\be
\theta_i (t)=\epsilon \varphi_i(t),
\ee
where $\varphi_i(t)$ are {\it bosonic} functions of the temporal variable and $\epsilon^2=0$. As follows from (\ref{SHam}), $\varphi_i(t)$ obey the linear differential equations
\bea\label{EXT}
&&
{\dot\varphi}_1=-\{\Lambda_{13},\Lambda_{23} \} \varphi_2-\{\Lambda_{12},\Lambda_{23} \} \varphi_3, \qquad {\dot\varphi}_2=\{\Lambda_{13},\Lambda_{23} \} \varphi_1-\{\Lambda_{12},\Lambda_{13} \} \varphi_3,
\nonumber\\[2pt]
&&
{\dot\varphi}_3=\{\Lambda_{12},\Lambda_{23} \} \varphi_1+\{\Lambda_{12},\Lambda_{13} \} \varphi_2.
\eea
Because a square of a Grassmann--odd number is zero, $\epsilon^2=0$, equations of motion for $(x_i,y_i)$ reduce to those of the original bosonic system governed by $H$ in (\ref{HHam})
\be\label{EEOM}
{\dot x}_i=\{x_i,H\}, \qquad  {\dot y}_i=\{y_i,H\}.
\ee

Thus, Eqs. (\ref{EXT}) describe an extension of (\ref{HHam}), (\ref{EEOM}) by three {\it bosonic} degrees of freedom $\varphi_i$. Because the new variables do not alter the dynamics of $(x_i,y_i)$, it suffices to establish integrability in the $\varphi_i$--sector. Two first integrals
\be\label{INT}
\Lambda_{12} \varphi_3+\Lambda_{13} \varphi_2+\Lambda_{23} \varphi_1, \qquad \varphi_1^2+\varphi_2^2+\varphi_3^2,
\ee
the first of which is obtained from the super--charge (\ref{sc}),
allow one to reduce (\ref{EXT}) to a single linear inhomogeneous first order differential equation, which can be easily integrated by conventional means.

As an illustration, let us consider the Hamiltonian (\ref{H2}), in which all $\Gamma_i$ are assumed positive. The building blocks
\be
\Lambda_{ij}=\sqrt{\Gamma_i \Gamma_j}  e^{{\left(x_i-x_j\right)}^2} e^{{\left(y_i-y_j \right)}^2},
\ee
with $i<j$, obey the structure relations
\bea\label{ALG}
&&
\{\Lambda_{12},\Lambda_{13} \}=\frac{4}{\Gamma_1} \left( (y_1-y_2)(x_1-x_3)-(y_1-y_3)(x_1-x_2)\right) \Lambda_{12} \Lambda_{13},
\nonumber\\[2pt]
&&
\{\Lambda_{12},\Lambda_{23} \}=-\frac{4}{\Gamma_2} \left( (y_1-y_2)(x_2-x_3)-(y_2-y_3)(x_1-x_2)\right) \Lambda_{12} \Lambda_{23},
\nonumber\\[2pt]
&&
\{\Lambda_{13},\Lambda_{23} \}=\frac{4}{\Gamma_3} \left( (y_1-y_3)(x_2-x_3)-(y_2-y_3)(x_1-x_3)\right) \Lambda_{13} \Lambda_{23},
\eea
which specify the equations of motion (\ref{EXT}) for the extra variables. Interestingly enough, the functions (no sum over repeated indices)
\be
(x_j-x_i)(y_k-y_i)-(x_k-x_i)(y_j-y_i),
\ee
which accompany the quadratic combinations of $\Lambda$ on the right hand sides of Eqs. (\ref{ALG}), coincide with the geometric variables $\Delta_{ijk}$ introduced in \cite{BBM}.

While $(x_i,y_i)$ follow the vortex orbits depicted in Fig. 3, each component of $\varphi_i$ undergoes a quasi--periodic oscillation (see Fig. 5). The tip of the vector $\varphi_i$ swings on a two--sphere (\ref{INT}) and it can be interpreted as a generalised spin vector \cite{AG}.

\begin{figure}[ht]
\begin{center}
\resizebox{0.5\textwidth}{!}{%
\includegraphics{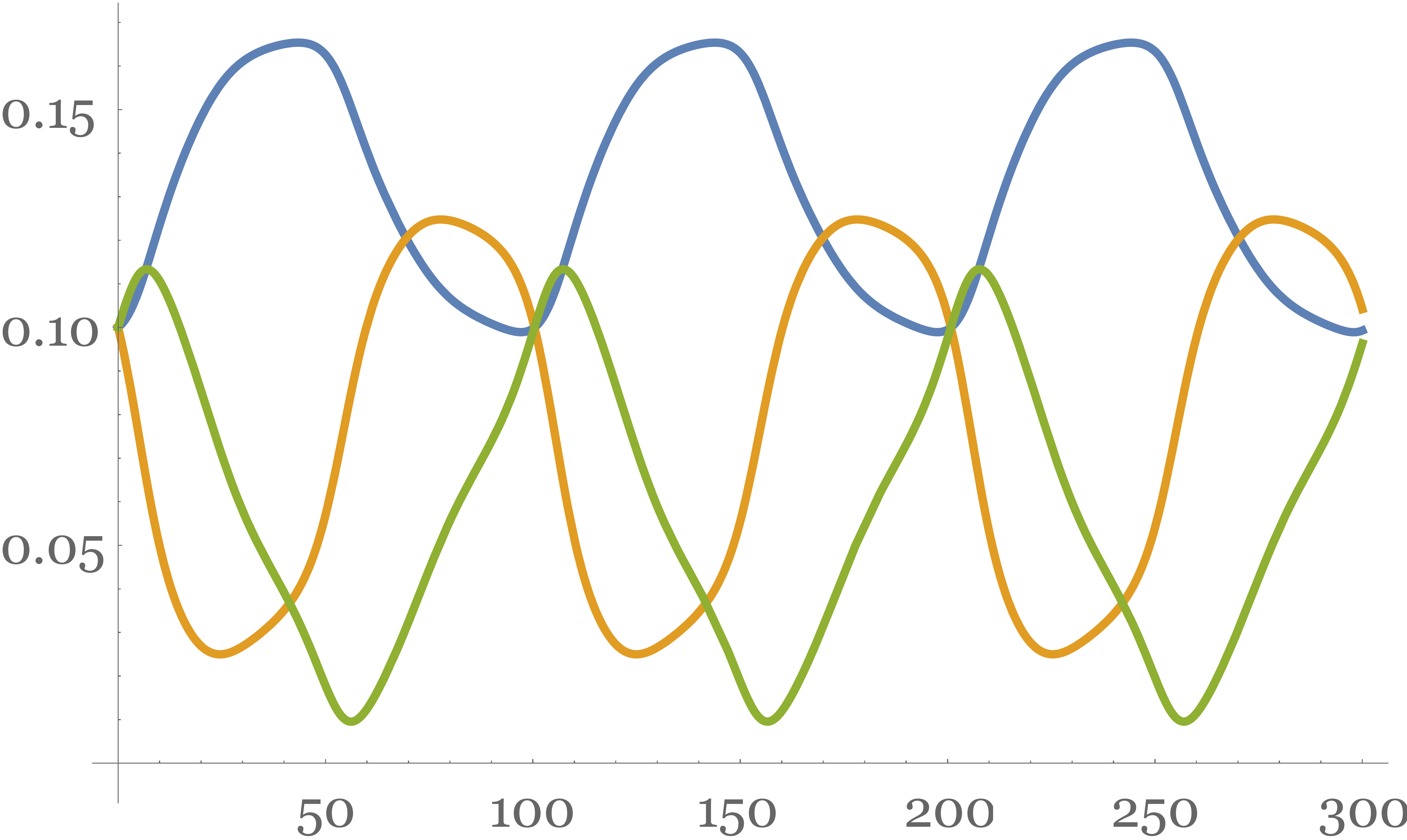}}\vskip-4mm
\caption{\small Plots $\varphi_1(t)$ (top), $\varphi_2(t)$ (middle), $\varphi_3(t)$ (bottom) corresponding to Eqs. (\ref{EXT}), for $\Gamma_1=0.1$, $\Gamma_2=0.3$, $\Gamma_3=0.5$, $x_1(0)=0.1$, $y_1 (0)=0.1$, $x_2 (0)=0.1$, $y_2 (0)=-0.1$, $x_3 (0)=-0.1$, $y_3 (0)=-0.1$, $\varphi_1(0)=0.1$, $\varphi_2(0)=0.1$, $\varphi_3(0)=0.1$, and $t \in [0,300]$.}
\label{fig3}
\end{center}
\end{figure}

\vspace{0.5cm}

\noindent
{\bf 5. Conclusion}\\

\noindent
To summarise, in this Letter integrable generalisations of a three--vortex system on a plane were studied.
First, restrictions on a form of a Hamiltonian, which follow from the $E(2)$--symmetry and the related integrability, were formulated. Then a few models, which rely upon root vectors of simple Lie algebras, were proposed.
Finally, it was demonstrated that the generalised systems governed by a positive definite  Hamiltonian can be extended by dynamical spin variables without destroying integrability.

Turning to possible further developments, it would be interesting to study in detail a link of the generalised models in this work to the nonrelativistic fluid mechanics. In particular, a possible modification of the Euler equations is worth studying. Some models in Sect. 3 bear resemblance to the Ruijsenaars--Schneider model \cite{RS}, which is known to be integrable for an arbitrary number of particles. It is interesting to explore whether some of the generalised models may admit extra integrals of motion in addition to those originating from the $E(2)$--symmetry. Long time behaviour of the generalised systems is worth studying as well.

\vspace{0.5cm}

\noindent{\bf Acknowledgements}\\

\noindent
This work was supported by the Tomsk Polytechnic University development program "Priority 2030", grant No 001-0000-2022.

\end{document}